\documentclass[12pt]{article}
\usepackage{graphicx}
\usepackage{color}
\usepackage{amsfonts}
\usepackage{latexsym}
\usepackage{amsmath,amssymb}
\usepackage{verbatim}
\usepackage{multirow,cite}

\usepackage[dvipdfm,hypertex]{hyperref}
\numberwithin{equation}{section}

\newcommand{\bea}{\begin{eqnarray}}
\newcommand{\eea}{\end{eqnarray}}
\newcommand{\be}{\begin{equation}}
\newcommand{\ee}{\end{equation}}

\def\nn{\nonumber}
\def\half{{\textstyle{1\over2}}}
\def\p{\partial}

\def\dltg{1+(\lambda \pi T_-Q)^2}
\def\cQ{\mathcal{Q}}
\begin{document}\begin{center}

{ \LARGE {\textsc{Warped AdS$_3$/Dipole-CFT Duality }}}

\vspace{0.8cm}

Wei Song and Andrew Strominger

\vspace{1cm}

\vspace{0.5cm}

{\it  Center for the Fundamental Laws of Nature, Harvard University,\\
Cambridge, MA 02138, USA}

\vspace{0.5cm}

\vspace{1.0cm}

\end{center}

\begin{abstract}
\noindent
\end{abstract}
  String theory contains solutions with $SL(2,\mathbb{R})_R\times U(1)_L$-invariant warped AdS$_3$ (WAdS$_3$) factors arising as continuous deformations of ordinary AdS$_3$ factors. We propose that some of these are holographically dual to the IR limits of nonlocal dipole-deformed 2D D-brane gauge theories, referred to as ``dipole CFTs". Neither the bulk nor boundary theories are currently well-understood, and consequences of the  proposed duality for both sides is investigated. The bulk entropy-area law  suggests that dipole CFTs  have (at large N) a high-energy density of states which does not depend on the deformation parameter. Putting the boundary theory on a spatial circle leads to closed timelike curves in the bulk, suggesting a relation of the latter to dipole-type nonlocality.

\pagestyle{empty}

\pagebreak
\setcounter{tocdepth}{2}
\tableofcontents
\section{Introduction}
The holographic AdS/CFT duality relating string theory on anti de Sitter space (AdS) to conformal field theory (CFT) is by now relatively well-understood. However AdS is a very special solution of string theory and a CFT is a very special kind of quantum field theory. It is important to try to generalize this duality in as many ways as possible.

In the last several years warped AdS$_3$ (WAdS$_3$) solutions of string  theory have emerged as interesting generalizations of AdS$_3$ solutions, see e.g.  \cite{waa,wab,wac,wad,wae,waf,wag,wah, Anninos:2009jt,Detournay:2010rh,wai}. They appear in the solution space of string theory as continuous deformations of ordinary AdS$_3$ which break the $SL(2,\mathbb{R})_R\times SL(2 ,\mathbb{R})_L$ isometry  down to $SL(2,\mathbb{R})_R\times U(1)_L$.   This continuous deformation however fundamentally changes the nature of the boundary, and apparently requires an as-yet-undiscovered holographic dictionary.\footnote{see \cite{Guica:2010sw,Ross:2011gu} for progress on related issues.}
Finding this dictionary is an important challenge which will likely yield new insights into the nature of holography.

In parallel with these developments, recent investigations have indicated an interesting class of nonlocal deformations of quantum  field theory known as dipole deformations \cite{Bergman:2000cw,Dasgupta:2000ry,Bergman:2001rw}.  These are simpler cousins  of quantum field theories on noncommutative spaces in which one of the spatial momenta is exchanged with a global conserved charge.  These dipole theories have a very rich and interesting structure about  which much remains to be fully understood.

In this note we will point out that these two types of structures, which lie at the edges of our understanding of bulk string theory and quantum field theory, are in some cases holographically dual. Our hope is that the relation between them will advance our understanding of both.

More specifically, the dipole deformed field theories dual to WAdS$_3$ solutions of string theory are the IR limits of dipole deformed 2D gauge theories, in some cases at finite temperature or chemical potential. We refer to these IR theories as dipole CFTs. They have a novel $SL(2,\mathbb{R})_R\times U(1)_L$ global symmetry. Field theories with novel scaling symmetries have also been a subject of much recent interest for condensed matter applications \cite{Hartnoll:2009sz}. The special cases of local field theories with
$SL(2,\mathbb{R})_R\times U(1)_L$ symmetries were considered in
\cite{Hofman:2011zj} and found to have an infinite symmetry enhancement.  The full symmetry group of dipole CFTs remains to be understood.

A further reason for interest in the WAdS$_3$/dipole CFT structure  is that it is a simpler cousin of the one appearing in the Kerr/CFT correspondence \cite{Guica:2008mu}.
Indeed the WAdS$_3$ geometry may well have  been seen in the sky
\cite{Sulentic:1998ik,McClintock:2006xd}
as fixed polar angle sections of extreme Kerr black holes.  Kerr/CFT has recently been embedded in string theory \cite{Guica:2010ej,Compere:2010uk,Song:2011ii}, in which context the holographic dual is in principle known. In trying to unravel this structure, we were led to the somewhat simpler WAdS$_3$/dipole CFT dualities studied in this paper. Many of the puzzling aspects of the Kerr/CFT correspondence (see \cite{Bredberg:2011hp} for a review)
already make an appearance in the  WAdS$_3$/dipole CFT context. Our hope is that investigations of WAdS$_3$/dipole CFT duality will ultimately shed insights into the structure of Kerr/CFT and astrophysical extreme black holes.

Some interesting features and new questions emerged in our analysis.  In section 2 we construct the geometry dual to the IR limit of the dipole deformed D1-D5 gauge theory on the Minkowski plane  with charges $Q_1=Q_5=Q$ and find a WAdS$_3$ factor with a null warping. In section 3, we construct the dipole deformation of the theory at finite temperature on the plane and find the warping becomes spacelike.  Two sets of boundary conditions are given which lead to asymptotic left or right
Virasoros with central charges $c_L =6Q^2$ or $c_R=6Q^2$, which do not depend on the dipole deformation parameter $\lambda$.  Moreover, the entropy density computed from the bulk solution using the area law agrees with that obtained by a naive application of the Cardy formula to the dipole CFT. These observations together suggest that dipole CFTs retain some of the characteristics of ordinary CFTs. Exactly which characteristics remains to be fully understood. In section 4 we move from the Minkowski plane to the cylinder by
periodic identification of the spatial coordinate of the dipole CFT. Interestingly we find that this generically leads, on the bulk side, to closed timelike curves (CTCs) near the WAdS$_3$ boundary, in the region  corresponding to scales shorter than the dipole length of the dipole CFT.  Perhaps this can lead to some insight on the nature of CTCs in gravity.
The one special case in which CTCs do not arise is when the periodic identification in the dipole CFT is null and aligned with the dipole direction itself.

Overlapping and compatible observations have been made in the recent paper \cite{ElShowk:2011cm}.
\section{Dipole deformation of the ground state}
We begin with the familiar near-horizon solution  of the supersymmetric D1-D5 string in its ground state
\bea\label{fty}
ds_6^2&=&Q({-2dx^+dx^-+dz^2\over z^2}+d\Omega^2_3)\label{planar}\\
F^{RR} &= &2{ {Q} } \left(\epsilon_3+*\epsilon_3
 \right)
 \\
e^{\Phi}&=&1
\eea
where $d\Omega_3$ is the metric on a unit 3-sphere, $\epsilon_3$ is the volume form on a unit 3-sphere, and $*$ denotes the  6 dimensional Hodge dual.
For simplicity we have taken the D1 and D5 charges both equal  to $Q$ and suppress the 4D K3 or $T^4$.  In our conventions $2G_N^{(6)}=\pi^2$.

The D1-D5 gauge theory admits a dipole deformation in which the ordinary product of fields is replaced by a nonlocal $\star$ product.There is an extensive literature on the dipole deformation of gauge theories beginning with \cite{Bergman:2000cw}, mostly in greater than 2 dimensions, and the construction of holographic duals by solution-generating techniques. We assume below some familiarity with this literature and will not fully review the subject here.   The nonlocal $\star$ product involves a choice of a constant dipole vector field $L^\mu$ in spacetime (in our case 1+1 dimensional) as well as a conserved $U(1)$ charge $q$.  For two functions $f$ and $g$ which are momentum and charge eigentstates one then has
\be f\star g =e^{iq_fL^\mu p_\mu^g-iq_gL^\mu p_\mu^f}fg. \ee
This is similar to the more familiar $\star$ product of noncummutative geometry, but with one momentum direction replaced by the $U(1)$ charge.

If the gauge theory has a known holographic dual $and$ the $U(1)$ charge in question is
a geometric Kaluza-Klein (KK) charge, then an algorithm is known \cite{Bergman:2001rw,Lunin:2005jy,Maldacena:2008wh}  for constructing the deformed dual geometry.\footnote{An algorithm should exist for any choice of $U(1)$ charge, but the general case has not yet been understood.}
It is called the ``TsT" transformation. First,  one compactifies and T-dualizes along the
dipole direction $L^\mu$. Next, one twists the boundary condition around the dipole circle by a rotation (or shift) along the $U(1)$ with magnitude proportional to the size of the deformation.
This twist (unless of course it is an integral multiple of $2\pi$) is not a symmetry of the theory and leads to a physically inequivalent solution. Finally one T-dualizes again along the dipole circle and decompactifies. As explained in \cite{Bergman:2000cw,Dasgupta:2000ry,Bergman:2001rw} the resulting new gravity solution is the dual of the dipole-deformed field theory.

In this paper we will apply this procedure  to configurations with only RR fields for which
the string frame metric can be written as \bea ds^2_d&=&ds^2_{d-2}+g_{11}(dx^{(1)}+A^{(1)})^2+g_{22}(dx^{(2)}+A^{(2)})^2\\ F^{NS}&=&0\eea
where $x^{(2)}$ is the dipole direction and $x^{(1)}$ is the $U(1)$ fiber.
The TsT transformation is equivalent to a T duality along $x^{(1)}$, followed by a shift $x^{(2)}\rightarrow x^{(2)}+\lambda x^{(1)}$ and then a second T duality along $x^{(1)}.$
After this TsT transformation, the solution becomes
\bea\label{tst}
d\tilde{s}^2_{d}&=&ds^2_{d-2}+e^{-2(\Phi-\tilde{\Phi})}[g_{11}(dx^{(1)}+A^{(1)})^2+g_{22}(dx^{(2)}+A^{(2)})^2]\\
e^{-2\tilde{\Phi}}&=&(1+\lambda^2g_{11}g_{22})e^{-2\Phi}
\nn\\ \tilde{F}^{RR}&=&F^{RR}
\nn\\ \tilde{F}^{NS}&=&-d\left[{\lambda g_{11}g_{22}\over 1+\lambda^2g_{11}g_{22}}(dx^{(1)}+A^{(1)})\wedge (dx^{(2)}+A^{(2)})\right]\nn
\eea
If the shift $\lambda=0$, since $T^2=1$, the transformation is trivial. Hence $\lambda$ parameterizes a continuous deformation of the D1-D5 gauge theory.

In this section we  consider the case where the dipole direction is null (along $x^-$) and the $U(1)$ charge is the KK charge associated to $\p_\psi$.
TsT transformation on a null circle can be obtained as a limit of a spacelike circle.
For a twist of size $\lambda$, limit of (\ref{tst}) under $g_{11}\rightarrow 0$ gives \bea
ds^2&=&Q({-2dx^+dx^-+dz^2\over z^2}-\lambda^2Q^2{{dx^+}^2\over 4z^4}+d\Omega^2_3)\\
F^{RR} &= &2Q \left(\epsilon_3+*\epsilon_3  \right)\nn\\
F^{NS}&=&\lambda Q^2\left(-{1\over2z^3}dz\wedge dx^+\wedge(d\psi+\cos\theta d\phi)+{dx^+\over 4z^2}\wedge \sin\theta d\theta\wedge d\phi \right)\label{planarTsT}\nn\\
e^{\Phi}&=&1\nn
\eea
This solution has $SL(2,\mathbb{R})_R\times SU(2)_L\times U(1)^2$ symmetry. The three dimensional part of the metric is null warped AdS$_3$ (for a discussion of this geometry see e.g. \cite{Anninos:2008fx}). The deformations are suppressed (become large) in the IR (UV) at $z\to \infty$ ($z\to 0$) in accord with the fact that
the corresponding operator is irrelevant. Nevertheless the space is homogeneous and all points are related by the action of an isometry.

\section{Dipole deformation at finite temperature}
\subsection{Deformed solution}
 In this subsection we construct the duals of the dipole deformed theory at finite temperature. In the undeformed theory, finite temperature solutions dual to the CFT on a circle are BTZ black holes and may be obtained as quotients of the ground state solutions. This construction of black holes as quotients is special to three dimensional theories with unbroken conformal symmetries. It does not work in higher dimensions. It also does not work once the dipole deformation is turned on, as smooth quotients of this type do not exist for null warped AdS$_3$ \cite{Anninos:2008fx}.

To construct the dual of the  finite temperature dipole CFT, one must first turn on a finite temperature and then deform. The finite temperature analog of the metric in (\ref{fty}) is
\be {ds^2\over Q}=\rho^2(-d\tau^2+(dx)^2)+\rho^2_0( \cosh \delta d\tau+\sinh \delta dx)^2+{d\rho^2\over\rho^2-\rho_0^2}+d^2\Omega_3 \ee
On the boundary $\rho=\infty$ the metric goes like $\rho^2(-d\tau^2+(dx)^2)$ so $\tau^\pm= \tau \pm x$ are null coordinates. However they both  become Rindler type coordinates related to $x^\pm$ in (\ref{fty})
 by \be \label{crd} x^\pm=\pm {\sqrt{\rho^2-\rho_0^2}\over\sqrt{2} \rho}e^{\pm 2\pi T_\pm \tau^\pm}.\ee Here the left and right Rindler temperatures are
\be T_{\pm}={1\over2\pi}\rho_0e^{\pm\delta}\ee
The presence of the $d\tau^{+2}$ and $d\tau^{-2}$ terms in the metric indicates a nonzero energy density arising from the  Rindler temperatures.
The area law gives a Bekenstein-Hawking entropy density per unit $x$
\be
{\delta S \over \delta x }= \pi Q^2(T_++T_-).\label{Sbtz}
\ee
This geometry  has an asymptotically flat extension
 describing a near-extremal temperature $(T_+,T_-)$ D1-D5 black string with $\tau^\pm$ asymptotically canonically normalized  longitudinal null coordinates \cite{Maldacena:1998bw}.
 The entropy can then be understood as the entropy per unit proper length along $x$ at temperatures $T_\pm$ .

The TsT construction involving T-duality along  $\tau^-$ rather than $x^-$ may then be interpreted as the dipole deformation of the finite temperature gauge theory.
 The 6D Einstein frame metric becomes
\be\label{ef}
{4d\tilde s^2\over Q\left(\dltg\right)^{1/2}}=-\omega_+\omega_-+\sigma_1^2+\sigma_2^2+{1\over \dltg}(\omega_3^2+\sigma_3^2)
\ee
where we have defined the $SU(2)_L\times SL(2,\mathbb{R})_R$ invariant one forms
\bea
\sigma_1 &=& \cos\psi d\theta + \sin\theta \sin\psi d\phi\\
\sigma_2 &=& -\sin\psi d\theta + \sin\theta \cos\psi d\phi\notag\\
\sigma_3 &=& d\psi + \cos\theta d\phi\notag\\
\omega_{\pm} &=& 4\pi T_+e^{\mp2\pi T_- \tau^-}{\rho\sqrt{(\rho^2-\rho _0^2)}\over\rho_0^2}\left(d\tau^+ \mp{2\pi T_- d\rho\over\rho(\rho^2-\rho_0^2)}\right)\notag\\
\omega_3 &=& 2\pi T_-d\tau^-+{(-2\rho^2+\rho_0^2) d\tau^+\over2\pi T_-} \ ,\notag
\eea
The new  6D dilaton is \be
e^{-2\tilde \Phi}
=\dltg\label{tstdlt}
\ee
The string frame metric is
$ ds^2_{string}\equiv e^{\tilde\Phi}ds^2$.

The three form fields becomes
\bea
{\tilde F}^{NS}&=&-{\lambda (\pi T_-)^2Q^2\over \dltg}({1\over2}\omega_+\wedge\omega_-\wedge\sigma_3+\sigma_1\wedge\sigma_2\wedge\omega_3)\\
{\tilde F}^{RR}&=&{Q\over4}({1\over2}\omega_+\wedge\omega_-\wedge\omega_3+\sigma_1\wedge\sigma_2\wedge\sigma_3)\label{tstf}
\eea
Note that the RR field is self dual, while the NS-NS field is anti-self dual.

The $SL(2,\mathbb{R})_R\times SL(2,\mathbb{R})_L$ portion of the isometry group is broken to  $SL(2,\mathbb{R})_R\times  U(1)_L$  with generators
\bea
\tilde{H}_\pm&=&{e^{\mp2\pi T_+\tau^+}\over 2\rho\sqrt{\rho^2-\rho_0^2}}\left({2\rho^2-\rho_0^2\over2\pi T_+}\p_++2\pi T_+\p_-\pm\rho(\rho^2-\rho_0^2)\p_\rho\right)\\
\tilde{H}_0&=&{1\over 2\pi T_+}\p_+
\\
H_L&=&-{\p_-\over2\pi T_-}
\eea
The $SL(2,\mathbb{R})_R$ generators are normalized so that they satisfy the standard  algebra $[\tilde{H_0},\tilde{H}_\pm]=\mp \tilde{H}_\pm,~[\tilde{H}_+,\tilde{H}_-]=2\tilde{H}_0.$

\subsection{Thermodynamics}
   In this subsection we consider the thermodynamics of the deformed black string geometry. We do not expect the temperatures $T_\pm$ to change because we carry out the deformation in the boundary Rindler coordinates (\ref{crd}). The entropy density (per unit $x$) is the  Einstein frame area of any  point of fixed $\tau^\pm$ divided by $2\pi^2=4G_N^{(6)}$. From (\ref{ef}) we see that the factors of $\dltg$ cancel and we regain  formula (\ref{Sbtz}):
\be \label{nnn}
{\delta S \over \delta x}= \pi Q^2(T_++T_-).
\ee
Interestingly, this relation is independent of $\lambda$.

We can also associate left and right central charges to the deformed geometry.  With appropriate choices of boundary conditions, the $SL(2,\mathbb{R})_R\times U(1)_L$ global isometry can be enhanced to either a left or a right conformal group.
 To see this, we first dimensionally reduce along $\psi$. Then the NS-NS field becomes a U(1) gauge field $A\propto \omega_3$. From the analysis of \cite{Compere:2009dp}, this U(1) gauge field does not contribute to the central charge. It may lead to a current algebra. We are mainly interested in the central charges in this paper, so we can just ignore the NS-NS field for that purpose. Then we can further reduce to warped $AdS_3$. Since the 6 dimensional spacetime is a direct product of warped $AdS_3$ and warped $S^3$, warped $S^3$ only contributes to the 3 dimensional Newton's constant. Similar metric and gauge field was analyzed in various papers, for example \cite{Compere:2007in,waf,ElShowk:2011cm}. Here we just lay out the main formulas. More details could be found in \cite{ElShowk:2011cm}. With the boundary condition  \bea  \label{bcL}\left(
                                        \begin{array}{ccc}
                                          h_{++}\sim \rho^4 & h_{+-}\sim \rho^{0} & h_{+\rho}\sim \rho^{-2} \\
                                           & h_{--}\sim\rho^{0} & h_{-\rho}\sim \rho^{-1} \\
                                             &  & h_{\rho\rho}\sim \rho^{-4} \\
                                         \end{array}
                                        \right)\,,\quad\quad\quad
                                         \cQ_{\tilde{H}_0}=0
                                        \eea
                                        where  $\cQ_{\tilde{H}_0}$ is the conserved charge associated with the vector $\cQ_{\tilde{H}_0}$,
the asymptotic symmetry generators are the right moving vector $\tilde{H}_0$ and the  left moving  vector fields
\bea \xi_{L}&=&-{1\over2}\p_-\epsilon_-(\tau^-)\rho\p_\rho+\epsilon_-(\tau^-)\p_-
\eea
The analysis of the of central charge associated to these generators  closely parallels \cite{Brown:1986nw,Compere:2007az,Guica:2008mu}.
The conserved charges $\cQ_{\xi_L}$ associated with the asymptotic Killing vectors $\xi_L$ are
symmetry generators on the phase space, and generate $\xi_L$  transformations via Dirac brackets. The Dirac bracket algebra of the conserved charges is the same as the Lie bracket algebra of the asymptotic Killing vectors, up to a central term:
\be
\delta_{\xi_L} \cQ_{\tilde{\xi}_L}\equiv \{\cQ_{\xi_L},\cQ_{\tilde{\xi}_L}\}_{D.B.}=\cQ_{[\xi_L,\tilde{\xi}_L]}+{1\over 8\pi G_3}\int _{\p\Sigma}k_{\xi_L}[\mathcal{L}_{\tilde{\xi}_L}g,g].
\ee
The central term $k_{\xi_L}[\mathcal{L}_{\tilde{\xi}_L}g,g]$ is a one-form, whose explicit expression can be found in \cite{Compere:2007az,Guica:2008mu}.
In this case, \be {1\over 8\pi G_3}\int _{\p\Sigma}k_{\xi_L}[\mathcal{L}_{\tilde{\xi}_L}\bar{g},\bar{g}]={Q^2\over4\pi}\int d\tau^-\epsilon_-'''(\tau^-)\tilde{\epsilon}_-(\tau^-)+\cdots\ee
where prime denotes derivative with respect to $\tau^-$ and $\cdots$ denotes terms with only one derivative.
The triple derivative term implies a left moving central charge
\be c_L=6Q^2\label{cL}.\ee

An alternate set of consistent boundary conditions gives a right-moving Virasoro.
Closely following the analysis of \cite{Castro:2009jf, Bredberg:2009pv, Hartman:2009nz, Castro:2010fd,Matsuo:2009sj},  these boundary conditions are
\be \left(
                                        \begin{array}{ccc}
                                          h_{++}\sim \rho^2 & h_{+-}\sim \rho^{0} & h_{+\rho}\sim \rho^{-1} \\
                                           & h_{--}\sim\rho^{0} & h_{-\rho}\sim \rho^{-1} \\
                                             &  & h_{\rho\rho}\sim \rho^{-4} \\
                                         \end{array}
                                        \right)\label{bcR}\ee
 supplemented by \bea \bar{g}^{\mu\nu}h_{\mu\nu}&\sim& \rho^{-3}\\
\p_-h_{+\rho}&\sim&\rho^{-2}
 \eea
The right moving Virasoro generators are found to be
\bea \xi_{R}&=&-{1\over2}\p_+\epsilon(\tau^+)\rho\p_\rho+\epsilon_+(\tau^+)(\p_+-{T_+\over T_-}\p_-)\\
\eea
The corresponding central term is
\be {1\over 8\pi G_3}\int _{\p\Sigma}k_{\tilde{\xi}_R}[\mathcal{L}_{\tilde{\xi}_R}\bar{g},\bar{g}]={Q^2\over4\pi}\int d\tau^+\epsilon_+'''(\tau^+)\tilde{\epsilon}_+(\tau^+)+\cdots\ee
giving the right moving central charge \be c_R=6Q^2\label{cR}\ee

So far no consistent boundary condition that allows both Virasoros has been found. Divergences of the charges appear if we natively relax (\ref{bcL}) or (\ref{bcR}) to allow both Virosoros. This puzzling situation is very
similar to one that has been discussed many times in the literature  in the context of the NHEK geometries of extreme Kerr \cite{Matsuo:2009sj,Matsuo:2009pg,Rasmussen:2009ix, Rasmussen:2010sa,Azeyanagi:2011zj}. We do not understand the significance of this phenomena. We hope that the present context, in which the microscopic dual is better understood, will prove a fruitful context in which to address the puzzle. We also note that recent investigations from a purely field theoretic perspective \cite{Hofman:2011zj} suggest the generic appearance of infinite dimensional symmetry groups in theories with global $SL(2,\mathbb{R})_R\times U(1)_L$ symmetries. It would be interesting to adapt the discussion of \cite{Hofman:2011zj} to the present context.

 Finally, we note that given the central charges and temperatures found above together with Cardy's formula gives the entropy density \be {\delta S \over \delta x}={1\over2\pi}{\pi^2\over3}(c_LT_-+c_RT_+)=\pi Q^2(T_++T_-).\ee
This agrees precisely with the  Bekenstein-Hawking result (\ref{nnn}). Since the dipole deformed theories are still incompletely understood this is not a microscopic derivation of the macroscopic area law. Rather, we should  regard it as a possible  indication that the
dipole deformation of a CFT does not alter high energy density of states.

\section{Putting the theory on a circle and CTCs}
So far we have studied 2D CFTs and their dipole deformations living  on the real line at finite temperature. In this section we investigate what happens when the real line is periodically identified to obtain a circle.

The identification\footnote{Of course we retain the identification $\psi\sim \psi +4\pi n$ required for for a smooth geometry at the poles of the sphere} corresponding to putting the boundary field theory on a circle is
\be x=\half(\tau^+-\tau^-)=x+2\pi mR\label{xcircle}\ee
The norm of the vector field $\p_x=2\pi T_+\tilde H_0+2\pi T_-H_L$ is
\be{4\over Q} | \p_x |^2={\rho^2\over \pi^2 T_-^2}(-\rho^2+4\pi^2T_+T_-) +{4\pi^2 \over \dltg}\left({\rho^2 \over 2 \pi ^2T_-}-T_++T_- \right)^2
\ee
near the horizon $\rho^2\to 4\pi^2T_+T_-$
\be {4\over Q}|\p_x|^2 \to {4\pi^2 \over \dltg}(T_++T_-)^2,\ee
which is spacelike. On the other hand for $\rho \to \infty$ we have
\be {4\over Q}|\p_x|^2 \to - {\rho^4\over \pi^2 T_-^2}{(\lambda \pi T_-Q)^2 \over \dltg}+{4\rho^2(1+(\lambda\pi T_-Q)^2{T_+\over T_-})\over\dltg}+\cdots,\ee
which is negative. Expanding around small $\lambda$, we see that the norm changes sign at around
\be \rho\sim \rho_c\equiv {2\over \lambda Q}\ee  which goes to infinity
for small $\lambda$ as expected. Hence the CTCs are associated to some funny nonlocal UV behavior of the conformal field theory.
Solutions of string theory with CTCs have appeared in a number of contexts \cite{ctca,ctcb,ctcc,ctcd,ctce,ctcf}, perhaps this example can shed some light on their meaning.

{
CTCs are present at infinity for generic, but not all, circle compactifications. An arbitrary Killing vector $\p_y\equiv v_1\p_t+v_2\p_x$ has norm
\bea {4\over Q}|\p_y^2|&=&-{(\lambda \pi T_-Q)^2(v_1+v_2)^2\over\pi^2T_-^2(\dltg)}\rho^2(\rho^2-4\pi^2T_+T_-)-{4(v_1^2-v_2^2)\over\dltg}\rho^2\\
&&+{4\pi^2\left(T_-(v_1-v_2)+{T_+}(v_1+v_2)\right)^2\over\dltg}\nn
\eea
Near the horizon $\rho^2\rightarrow4\pi^2T_+T_-$,  it goes to
\bea {4\over Q}|\p_y^2|&\sim&{4\pi^2\left(T_-(v_1-v_2)-{T_+}(v_1+v_2)\right)^2\over\dltg}
\eea
which is always positive.
If $v_1\neq v_2$,
the norm near the infinity $\rho\rightarrow \infty$ goes to
\bea {4\over Q}|\p_y^2|&\sim&-{(\lambda \pi T_-Q)^2(v_1+v_2)^2\over\pi^2T_-^2(\dltg)}\rho^4
\eea
which is negative.
For the vector with $v_1=-v_2,$ the norm is a positive constant,
\be {4\over Q}|\p_y^2|={16\pi^2 T_-^2v_1^2\over\dltg}\ee
Therefore we see that there is no everywhere timelike Killing vector. Furthermore if the field theory is put on an arbitrary circle, there are almost always closed timelike curves near infinity. The exception is the self-dual case when the circle is generated by  \be \p_y=\p_-=2\pi T_-H_L\,.\ee
}
Theory on the above circle and theory on (\ref{xcircle}) are related by an infinite boost \cite{Balasubramanian:2009bg}.

\section*{Acknowledgements}
We are grateful to Geoffrey Comp$\grave{\hbox{e}}$re, Sheer El-Showk, Alessandra Gnecchi, Monica Guica,
Josh Lapan and Juan Maldacena for useful conservations. This work was
supported in part by DOE grant DE-FG02-91ER40654 and the Harvard Society of Fellows.


\begin{thebibliography}{99}

\bibitem{waa}
  D.~Israel, C.~Kounnas, D.~Orlando, P.~M.~Petropoulos,
  ``Electric/magnetic deformations of S**3 and AdS(3), and geometric cosets,''
  Fortsch.\ Phys.\  {\bf 53}, 73-104 (2005).
  [hep-th/0405213].

\bibitem{wab}
  S.~Detournay, D.~Orlando, P.~M.~Petropoulos, P.~Spindel,
  ``Three-dimensional black holes from deformed anti-de Sitter,''
  JHEP {\bf 0507}, 072 (2005).
  [hep-th/0504231].

 \bibitem{wac}
  D.~Anninos, W.~Li, M.~Padi, W.~Song, A.~Strominger,
  ``Warped AdS(3) Black Holes,''
  JHEP {\bf 0903}, 130 (2009).
  [arXiv:0807.3040 [hep-th]].
\bibitem{wad}
  D.~Anninos,
  ``Hopfing and Puffing Warped Anti-de Sitter Space,''
  JHEP {\bf 0909}, 075 (2009).
  [arXiv:0809.2433 [hep-th]].

 \bibitem{wae}
  D.~Anninos, M.~Esole, M.~Guica,
  ``Stability of warped AdS(3) vacua of topologically massive gravity,''
  JHEP {\bf 0910}, 083 (2009).
  [arXiv:0905.2612 [hep-th]].

  \bibitem{waf}
  G.~Compere, S.~Detournay,
  ``Boundary conditions for spacelike and timelike warped AdS3 spaces in topologically massive gravity,''
  JHEP {\bf 0908}, 092 (2009).
  [arXiv:0906.1243 [hep-th]].


\bibitem{Anninos:2009jt}
  D.~Anninos,
  ``Sailing from Warped AdS(3) to Warped dS(3) in Topologically Massive Gravity,''
  JHEP {\bf 1002}, 046 (2010).
  [arXiv:0906.1819 [hep-th]].

 \bibitem{wag}
  B.~Chen, B.~Ning, Z.~-b.~Xu,
  ``Real-time correlators in warped AdS/CFT correspondence,''
  JHEP {\bf 1002}, 031 (2010).
  [arXiv:0911.0167 [hep-th]].


  \bibitem{wah}
  D.~Orlando, L.~I.~Uruchurtu,
  ``Warped anti-de Sitter spaces from brane intersections in type II string theory,''
  JHEP {\bf 1006}, 049 (2010).
  [arXiv:1003.0712 [hep-th]].


\bibitem{Detournay:2010rh}
  S.~Detournay, D.~Israel, J.~M.~Lapan, M.~Romo,
  ``String Theory on Warped $AdS_{3}$ and Virasoro Resonances,''
  JHEP {\bf 1101}, 030 (2011).
  [arXiv:1007.2781 [hep-th]].


  \bibitem{wai}
  M.~Henneaux, C.~Martinez, R.~Troncoso,
  ``Asymptotically warped anti-de Sitter spacetimes in topologically massive gravity,''
   [arXiv:1108.2841 [hep-th]].


\bibitem{Guica:2010sw}
  M.~Guica, K.~Skenderis, M.~Taylor, B.~C.~van Rees,
  ``Holography for Schrodinger backgrounds,''
  JHEP {\bf 1102}, 056 (2011).
  [arXiv:1008.1991 [hep-th]].

\bibitem{Ross:2011gu}
  S.~F.~Ross,
  ``Holography for asymptotically locally Lifshitz spacetimes,''
  [arXiv:1107.4451 [hep-th]].


\bibitem{Bergman:2000cw}
  A.~Bergman and O.~J.~Ganor,
  ``Dipoles, twists and noncommutative gauge theory,''
  JHEP {\bf 0010}, 018 (2000)
  [arXiv:hep-th/0008030].

\bibitem{Dasgupta:2000ry}
  K.~Dasgupta, O.~J.~Ganor and G.~Rajesh,
  ``Vector deformations of N=4 superYang-Mills theory, pinned branes, and
  arched strings,''
  JHEP {\bf 0104}, 034 (2001)
  [arXiv:hep-th/0010072].

\bibitem{Bergman:2001rw}
  A.~Bergman, K.~Dasgupta, O.~J.~Ganor, J.~L.~Karczmarek, G.~Rajesh,
  ``Nonlocal field theories and their gravity duals,''
  Phys.\ Rev.\  {\bf D65}, 066005 (2002).
  [hep-th/0103090].

\bibitem{Hartnoll:2009sz}
  S.~A.~Hartnoll,
  ``Lectures on holographic methods for condensed matter physics,''
  Class.\ Quant.\ Grav.\  {\bf 26}, 224002 (2009).
  [arXiv:0903.3246 [hep-th]].

\bibitem{Hofman:2011zj}
  D.~M.~Hofman, A.~Strominger,
  ``Chiral Scale and Conformal Invariance in 2D Quantum Field Theory,''

\bibitem{Guica:2008mu}
  M.~Guica, T.~Hartman, W.~Song, A.~Strominger,
  ``The Kerr/CFT Correspondence,''
  Phys.\ Rev.\  {\bf D80}, 124008 (2009).
  [arXiv:0809.4266 [hep-th]].

\bibitem{Sulentic:1998ik}
  J.~W.~Sulentic, P.~Marziani, M.~Calvani,
  ``Disk models for mcg-06-30-15: the variability challenge,''
  [astro-ph/9802305].

\bibitem{McClintock:2006xd}
  J.~E.~McClintock, R.~Shafee, R.~Narayan, R.~A.~Remillard, S.~W.~Davis, L.~-X.~Li,
  ``The Spin of the Near-Extreme Kerr Black Hole GRS 1915+105,''
  Astrophys.\ J.\  {\bf 652}, 518-539 (2006).
  [astro-ph/0606076].

\bibitem{Guica:2010ej}
  M.~Guica and A.~Strominger,
  ``Microscopic Realization of the Kerr/CFT Correspondence,''
  JHEP {\bf 1102}, 010 (2011)
  [arXiv:1009.5039 [hep-th]].

\bibitem{Compere:2010uk}
  G.~Compere, W.~Song, A.~Virmani,
 ``Microscopics of Extremal Kerr from Spinning M5 Branes,''
  [arXiv:1010.0685 [hep-th]].
\bibitem{Song:2011ii}
  W.~Song, A.~Strominger,
  ``D-brane Construction of the 5D NHEK Dual,''
  [arXiv:1105.0431 [hep-th]].

\bibitem{Bredberg:2011hp}
  I.~Bredberg, C.~Keeler, V.~Lysov, A.~Strominger,
  ``Cargese Lectures on the Kerr/CFT Correspondence,''
  Nucl.\ Phys.\ Proc.\ Suppl.\  {\bf 216}, 194-210 (2011).
  [arXiv:1103.2355 [hep-th]].



\bibitem{ElShowk:2011cm}
  S.~El-Showk and M.~Guica,
  ``Kerr/CFT, dipole theories and nonrelativistic CFTs,''
  arXiv:1108.6091 [hep-th].

\bibitem{Lunin:2005jy}
  O.~Lunin and J.~M.~Maldacena,
``Deforming field theories with U(1) x U(1) global symmetry and their gravity
  JHEP {\bf 0505}, 033 (2005)
  [arXiv:hep-th/0502086].

\bibitem{Maldacena:2008wh}
  J.~Maldacena, D.~Martelli, Y.~Tachikawa,
  ``Comments on string theory backgrounds with non-relativistic conformal symmetry,''
  JHEP {\bf 0810}, 072 (2008).
  [arXiv:0807.1100 [hep-th]].


\bibitem{Anninos:2008fx}
  D.~Anninos, W.~Li, M.~Padi, W.~Song, A.~Strominger,
``Warped AdS(3) Black Holes,''
  JHEP {\bf 0903}, 130 (2009).
  [arXiv:0807.3040 [hep-th]].

\bibitem{Maldacena:1998bw}
  J.~M.~Maldacena, A.~Strominger,
  ``AdS(3) black holes and a stringy exclusion principle,''
  JHEP {\bf 9812}, 005 (1998).
  [hep-th/9804085].



\bibitem{Compere:2009dp}
  G.~Compere, K.~Murata, T.~Nishioka,
  ``Central Charges in Extreme Black Hole/CFT Correspondence,''
  JHEP {\bf 0905}, 077 (2009).
  [arXiv:0902.1001 [hep-th]].

\bibitem{Compere:2007in}
  G.~Compere, S.~Detournay,
  ``Centrally extended symmetry algebra of asymptotically Godel spacetimes,''
  JHEP {\bf 0703}, 098 (2007).
  [hep-th/0701039].

\bibitem{Brown:1986nw}
  J.~D.~Brown, M.~Henneaux,
  ``Central Charges in the Canonical Realization of Asymptotic Symmetries: An Example from Three-Dimensional Gravity,''
  Commun.\ Math.\ Phys.\  {\bf 104}, 207-226 (1986).

\bibitem{Compere:2007az}
  G.~Compere,
  ``Symmetries and conservation laws in Lagrangian gauge theories with applications to the mechanics of black holes and to gravity in three dimensions,''

  [arXiv:0708.3153 [hep-th]].

\bibitem{Castro:2009jf}
  A.~Castro and F.~Larsen,
  ``Near Extremal Kerr Entropy from AdS(2) Quantum Gravity,''
  JHEP {\bf 0912}, 037 (2009)
  [arXiv:0908.1121 [hep-th]].

\bibitem{Bredberg:2009pv}
  I.~Bredberg, T.~Hartman, W.~Song and A.~Strominger,
  ``Black Hole Superradiance From Kerr/CFT,''
  JHEP {\bf 1004}, 019 (2010)
  [arXiv:0907.3477 [hep-th]].

\bibitem{Hartman:2009nz}
  T.~Hartman, W.~Song and A.~Strominger,
  ``Holographic Derivation of Kerr-Newman Scattering Amplitudes for General
  Charge and Spin,''
  JHEP {\bf 1003}, 118 (2010)
  [arXiv:0908.3909 [hep-th]].

\bibitem{Castro:2010fd}
  A.~Castro, A.~Maloney and A.~Strominger,
  ``Hidden Conformal Symmetry of the Kerr Black Hole,''
  Phys.\ Rev.\  D {\bf 82}, 024008 (2010)
  [arXiv:1004.0996 [hep-th]].

\bibitem{Matsuo:2009sj}
  Y.~Matsuo, T.~Tsukioka, C.~-M.~Yoo,
``Another Realization of Kerr/CFT Correspondence,''
  Nucl.\ Phys.\  {\bf B825}, 231-241 (2010).
  [arXiv:0907.0303 [hep-th]].

\bibitem{Matsuo:2009pg}
  Y.~Matsuo, T.~Tsukioka and C.~M.~Yoo,
  ``Yet Another Realization of Kerr/CFT Correspondence,''
  Europhys.\ Lett.\  {\bf 89}, 60001 (2010)
  [arXiv:0907.4272 [hep-th]].

\bibitem{Rasmussen:2009ix}
  J.~Rasmussen,
``Isometry-preserving boundary conditions in the Kerr/CFT correspondence,''
  Int.\ J.\ Mod.\ Phys.\  A {\bf 25}, 1597 (2010)
  [arXiv:0908.0184 [hep-th]].



\bibitem{Rasmussen:2010sa}
  J.~Rasmussen,
  ``A near-NHEK/CFT correspondence,''
  Int.\ J.\ Mod.\ Phys.\  A {\bf 25}, 5517 (2010)
  [arXiv:1004.4773 [hep-th]].

\bibitem{Azeyanagi:2011zj}
  T.~Azeyanagi, N.~Ogawa and S.~Terashima,
  ``On Non-Chiral Extension of Kerr/CFT,''
  arXiv:1102.3423 [hep-th].


\bibitem{ctca}
  G.~Compere, S.~Detournay, M.~Romo,
  ``Supersymmetric Godel and warped black holes in string theory,''
  Phys.\ Rev.\  {\bf D78}, 104030 (2008).
  [arXiv:0808.1912 [hep-th]].

\bibitem{ctcb}
  T.~S.~Levi, J.~Raeymaekers, D.~Van den Bleeken, W.~Van Herck and B.~Vercnocke,
``Godel space from wrapped M2-branes,''
  JHEP {\bf 1001}, 082 (2010)
  [arXiv:0909.4081 [hep-th]].

\bibitem{ctcc}
  J.~B.~Griffiths, N.~O.~Santos,
  ``A rotating cylinder in an asymptotically locally anti-de Sitter background,''
  Class.\ Quant.\ Grav.\  {\bf 27}, 125004 (2010).
  [arXiv:1003.3623 [gr-qc]].

\bibitem{ctcd}
  J.~Raeymaekers, D.~Van den Bleeken, B.~Vercnocke,
  ``Chronology protection and the stringy exclusion principle,''
  JHEP {\bf 1104}, 037 (2011).
  [arXiv:1011.5693 [hep-th]].

\bibitem{ctce}
  B.~Vercnocke,
  ``Hidden Structures of Black Holes,''
  [arXiv:1011.6384 [hep-th]].


\bibitem{ctcf}
  J.~Raeymaekers,
  ``Chronology protection in stationary three-dimensional spacetimes,''
   [arXiv:1106.5098 [hep-th]].




\bibitem{Balasubramanian:2009bg}
  V.~Balasubramanian, J.~de Boer, M.~M.~Sheikh-Jabbari, J.~Simon,
  ``What is a chiral 2d CFT? And what does it have to do with extremal black holes?,''
  JHEP {\bf 1002}, 017 (2010).
  [arXiv:0906.3272 [hep-th]].


















\end{thebibliography}
\end{document}